\documentstyle[prl,aps,epsfig,floats]{revtex}

\newcommand{\bib}{\bibitem}
\newcommand\bea{\begin{eqnarray}}
\newcommand\eea{\end{eqnarray}}
\newcommand\beq{\begin{equation}}
\newcommand\eeq{\end{equation}}
\newcommand\non{\nonumber}
\newcommand{\ua}{\uparrow}
\newcommand{\da}{\downarrow}
\newcommand{\al}{\alpha}
\newcommand{\be}{\beta}
\newcommand{\ep}{\epsilon}
\newcommand{\ga}{\gamma}
\newcommand{\om}{\omega}

\begin{document}

\draft

\textheight=24cm

\title{\Large \bf Equation of motion approach to non-adiabatic quantum charge
pumping}
\author{\bf Amit Agarwal and Diptiman Sen}
\address{\it Centre for High Energy Physics, Indian Institute of Science,
Bangalore 560012, India}

\date{\today}
\maketitle

\begin{abstract}
We use the equations of motion of non-interacting electrons in a 
one-dimensional system to numerically study different aspects of charge 
pumping. We study the effects of the pumping frequency, amplitude, band 
filling and finite bias on the charge pumped per cycle, and the Fourier 
transforms of the charge and energy currents in the leads. Our method works 
for all values of parameters, and gives the complete time-dependences of the 
current and charge at any site of the system. Our results agree with Floquet 
and adiabatic scattering theory where these are applicable, and provides 
support for a mechanism proposed elsewhere for charge pumping by a traveling 
potential wave. For non-adiabatic and strong pumping, the charge and energy 
currents are found to have a marked asymmetry between the two leads, and 
pumping can work even against a substantial bias.
\end{abstract}
\vskip .5 true cm

\pacs{~~ PACS number: ~73.23.-b, ~72.10.Bg, ~73.63.Nm}
\vskip.5pc

\section{Introduction}

The idea that periodically oscillating potentials applied at certain sites 
of a system can transfer a net charge per cycle between two leads (which are 
at the same chemical potential) has been studied for several years, both 
theoretically [1-18] and experimentally \cite{switkes,talyanskii}. Theoretical
studies have used adiabatic scattering theory \cite{avron,entin1}, Floquet 
scattering theory \cite{moskalets,kim} and variations of the non-equilibrium 
Green function (NEGF) formalism \cite{wang,arrachea,torres}. While Floquet 
scattering theory and the NEGF formalism work for potentials oscillating with 
any frequency, adiabatic theory works only for low frequencies. All these 
methods provide expressions for the charge transferred per cycle. However, it 
is not easy to obtain from these methods the current and charge at any site 
as a function of time. Detailed information like this may shed light on the 
mechanism of charge pumping, for instance, by a traveling potential
wave which has been observed in several experiments \cite{talyanskii}. Further,
the effects of a finite bias between the leads are not easy to study 
analytically unless the pumping is adiabatic \cite{entin2}. In this paper, we
present a numerical method for obtaining all this information for a system of 
non-interacting electrons. Our method is based on solving the equation of 
motion (EOM) of the density matrix of the system \cite{dhar1,dhar2,bushong}. 
(For time-independent Hamiltonians, this method reproduces the results obtained
by the NEGF formalism \cite{datta}). We will see that the time-dependences of 
the currents and charges depend significantly on the amplitude and frequency 
of the pumping. We will show that there is a marked asymmetry in the Fourier 
transforms of the charge and heat currents in the two leads if the pumping 
amplitude and frequency are large. We will also show that pumping can work 
even if there is a substantial bias opposing it.

As a simple model for studying charge pumping, we consider a one-dimensional 
system consisting of two semi-infinite leads $a=L,R$ (denoting left and right)
and a finite region $W$ (wire) lying between the two. We will model all three
regions by lattices with electrons governed by a one-channel tight-binding 
Hamiltonian with the same hopping amplitude $-\ga$ on all bonds, namely,
\beq
{\hat H}_0 ~=~ -\ga ~\sum_{n=1}^{N-1} ~(~ c_{n+1}^\dag c_n ~+~ c_n^\dag 
c_{n+1} )~,
\label{h0}
\eeq
where $N$ is the total number of sites. We will consider spinless electrons
here; the current of spin-1/2 electrons is simply twice that of spinless 
electrons for non-interacting electrons. The dispersion of the electrons in 
the leads is $E_k = -2\ga \cos k$, where $k$ lies in the range $[-\pi ,\pi]$.
(We are setting the Planck constant $\hbar$ and the lattice spacing equal to 
unity). The two leads are assumed to have the same chemical potential $\mu$ 
and temperature $T$. Time-dependent potentials will be applied to some sites 
of the wire; that part of the Hamiltonian is given by 
\beq
{\hat V} (t) ~=~ \sum_n ~V_n (t) ~c_n^\dag c_n ~, \quad {\rm where} \quad 
V_n (t) ~=~ a_n ~\cos (\om t + \phi_n) ~.
\label{pot}
\eeq
The sites with these potentials will be collectively called the scattering 
region.

In Secs. II and III, we will describe how Floquet scattering theory and 
adiabatic scattering theory respectively can be used to study charge pumping 
in the above model. In Sec. IV, we will describe the EOM method for numerically
studying various quantities of interest. Numerical results will be presented 
in Sec. V. In Sec. VI, we will summarize our results and point out some 
problems for future studies.

\section{Floquet scattering theory} 

Briefly, Floquet scattering theory works as follows \cite{moskalets,kim}. The
incoming electrons of energy $E_0$ gain or lose energy in quanta of $\om$ on 
interacting with the scattering region. Hence, the outgoing states are 
characterized by energies $E_p = E_0 + p \om$, where $p=0,\pm 1,\pm2, \cdots$;
the energies with $p \ne 0$ are called the Floquet side bands. 
The effect of the scattering region can be described by a Floquet
scattering matrix $S_{\al \be} (E_p, E_0)$, which is the amplitude for
an electron with energy $E_0$ entering through lead $\be$ to leave with
energy $E_p$ through lead $\al$. In the leads, 
the propagating modes have energies lying within the bandwidth $[-2\ga ,2 
\ga]$; only these modes can contribute to charge pumping. States with energies
lying outside the bandwidth have wave functions which decay exponentially into
the leads and hence do not contribute to charge transfer. The wave function 
of an electron coming from the left lead with an energy $E_0$ 
and wavenumber $k_0$ (with $E_0 = - 2 \ga \cos k_0$) is given by
\beq
\psi (n) ~=~ e^{i(k_0n - E_0 t)} ~+~ \sum_p ~r_p ~e^{i(-k_p n - E_p t)}~,
\eeq
at a site $n$ far to the left of the scattering region, and
\beq
\psi (n) ~=~ \sum_p ~t_p ~e^{i(k_p n - E_p t)} ~,
\eeq
far to the right of the scattering region, where $E_p = - 2 \ga \cos k_p$, and
the sums over $p$ run over values such that $E_p$ lies within the bandwidth of
the leads. The quantities $r_p$ and $t_p$ denote reflection and transmission 
amplitudes in the different side bands; they respectively denote the elements 
$S_{LL} (E_p, E_0)$ and $S_{RL} (E_p, E_0)$ of the Floquet scattering
matrix, where $L$ and $R$ denote the left and right leads. 
Similarly, the wave function of an electron coming 
from the right lead with an energy $E_0$ and wavenumber $k_0$ is given by
\beq
\psi (n) ~=~ e^{i(-k_0n - E_0 t)} ~+~ \sum_p ~\bar{r}_p ~e^{i(k_p n - E_p t)}~,
\eeq
far to the right of the scattering region, and
\beq
\psi (n) ~=~ \sum_p ~\bar{t}_p ~e^{i(-k_p n - E_p t)} ~,
\eeq
far to the left of the scattering region. The reflection and transmission 
amplitudes are found by writing down the wave functions in the scattering 
region, and matching coefficients of terms having the same time dependence 
($e^{\pm i E_p t}$) in the Schr\"odinger equation at different sites. If the 
oscillating potentials are weak, the reflection and transmission amplitudes 
decrease rapidly as $|p|$ increases; at first order in the potentials, only 
$p = \pm 1$ contribute. The current in, say, the right lead is then given by
\bea
& & I_R =~ q ~\int_{-2\ga}^{2\ga} ~\frac{dE_0}{2\pi} ~[~ \frac{v_1}{v_0} ~
(|t_1|^2 + |\bar{r}_1|^2 ) ~\{ f(E_0, \mu, T) - f(E_1, \mu, T) \} \non \\
& & ~~~~~~~~~~~~~~~~~~~~~~~~~~~+~ \frac{v_{-1}}{v_0} ~(|t_{-1}|^2 + 
|\bar{r}_{-1}|^2 ) ~\{ (f(E_0, \mu, T) - f(E_{-1}, \mu, T) \} ~] ~,
\label{ir}
\eea
where $f (E, \mu ,T) ~=~ 1/[e^{(E - \mu)/k_B T} + 1]$ is the Fermi function,
and $v_p = 2 \ga \sin k_p$ is the velocity. In the limit $\om \to 0$, we have 
$f(E_{\pm 1}, \mu, T) - f(E_0, \mu, T) = \pm ~\om \partial f(E_0, \mu, T) /
\partial E_0$, and $v_p/v_0 \to 1$. Finally, at zero temperature, $\partial 
f(E_0, \mu, 0) /\partial E_0 = - \delta (E_0 - \mu)$.

\section{Adiabatic scattering theory} 

In the limit of the pumping frequency tending to zero, the charge 
transport can be related to the `frozen' scattering matrix $S$ 
\cite{buttiker,brouwer1,avron}. The average pumped current in this limit is
found to be proportional to the frequency $\om$, and therefore the charge
pumped per cycle (of time period $2\pi /\om$) is independent of $\om$. 
If the scattering region is connected to 
leads which are at the same chemical potential and zero temperature, the 
infinitesimal charge flowing from that region to the $a$-th lead is given by
\beq 
dQ_a~=~ \frac{iq}{2\pi} ~(dS~S^\dag)_{aa} ~, 
\label{btp}
\eeq
where $q$ is the electron charge, and the `frozen' $S$-matrix is evaluated at 
the Fermi energy $E_F = - 2 \ga \cos k_F = \mu$. Eq. (\ref{btp}) can be used 
to relate adiabatic scattering theory to a geometric description of charge 
transport \cite{brouwer1}. For weak oscillating potentials $V_n$ as given in 
Eq. (\ref{pot}), one finds that the charge entering lead $a$ per cycle is 
given by
\beq 
\Delta Q_a = -~ \frac{q}{\pi} ~\sum_{n>m} ~Im~ \Bigl[ ~\Bigl( \frac{\partial 
S}{\partial V_n} \frac{\partial S^\dag}{\partial V_m} \Bigr)_{\{V_m \} =0}~
\Bigr]_{aa} ~
\oint ~dV_n ~V_m~ ,
\label{brou}
\eeq
where the integral is done over one cycle of the oscillation. Eq. (\ref{brou})
is a generalization of the formula in Ref. \cite{brouwer1} and can be derived
from Eq. (\ref{btp}) by Taylor expanding the scattering matrix to first order,
$S(V_m) = S(0) + \sum_n V_n (\partial S /\partial V_n)_{\{V_m \} =0}$. In the 
case of two weak oscillating potentials $V_1(t)$ and $V_2 (t)$, Eq. 
(\ref{brou}) shows that the pumped charge is proportional to the area 
in the space of the parameters $(V_1,V_2)$.

\section{Equation of motion method} 

For a system with a finite number of sites, we study the time evolution as 
follows. The density matrix of the system evolves according to the EOM
\beq
\hat{\rho}(t+dt) ~=~ e^{-i\hat{H} (t) dt} ~\hat{\rho}(t) ~
e^{i\hat{H} (t) dt} ~, 
\label{rhot}
\eeq
where $\hat{H} (t) = {\hat H}_0 + {\hat V} (t)$ is given in Eqs. 
(\ref{h0}-\ref{pot}). The current across any bond is then obtained by taking 
the trace of the appropriate current operator with $\hat{\rho}$. The current 
operator on the bond from site $n$ to site $n+1$ and its expectation value 
at time $t$ are given by 
\bea
\hat{j}_{n+1/2} &=& iq \ga ~(c_{n+1}^\dag c_n - c_n^\dag c_{n+1})~, \non \\
{\rm and} \quad j_{n+1/2} (t) &=& Tr (\hat{\rho} (t) \hat{j}_{n+1/2} ) ~=~ 
iq \ga ~[ \hat{\rho}_{n,n+1}(t) - \hat{\rho}_{n+1,n}(t) ]. \non \\
& &
\label{jnt}
\eea
The charge transferred between the right and left leads $R$ and $L$ between
two times can be found either by integrating the above expression in time, 
or by taking the operator
\beq
\Delta {\hat Q} ~=~ \frac{q}{2} ~\bigl[ ~\sum_{n \in R} ~c_n^\dag c_n ~-~ 
\sum_{n \in L}~ c_n^\dag c_n ~\bigr] ~,
\label{dqt}
\eeq
and computing $Tr~ (\hat{\rho} (t) \Delta \hat{Q})$ at the two times; these 
methods give the same result for the charge transferred in a cycle. 
Similarly, one can compute the energy current across any site $n$; the 
corresponding operator and its expectation value are given by
\bea
\hat{e}_n &=& -i \ga^2 ~(c_{n+1}^\dag c_{n-1} - c_{n-1}^\dag c_{n+1})~, \non \\
{\rm and} \quad e_n(t) &=& Tr (\hat{\rho} (t) \hat{e}_n) ~=~ -i \ga^2 ~[ 
\hat{\rho}_{n-1,n+1}(t) - \hat{\rho}_{n+1,n-1}(t) ], \non \\
& &
\label{ent}
\eea
in a region where the on-site potential $V_n$ is zero \cite{dhar1}. At zero 
temperature, the charge current is carried by electrons whose average energy
is given by $E_F = \mu$; the difference between the energy current $e$ and the
charge current $j$ multiplied by $\mu$ gives the heat current 
\cite{avron,moskalets}. We define the heat current at site $n$ to be 
$e_n - (\mu /2) (j_{n-1/2} + j_{n+1/2})$.

In all our calculations, we take the left and right leads to have $N_L$ sites
each and the wire in the middle to have $N_W$ sites; the total number of sites
is $N=2N_L +N_W$. We set the hopping amplitude $\ga=1$. For calculations in 
which the chemical potential is the same in the two leads, we take the density
matrix at time $t=0$ to be given by that of a single system governed by the 
Hamiltonian $H_0$ in Eq. (\ref{h0}) with $N$ sites, chemical potential $\mu$ 
and temperature $T$ (which we will take to be zero). If $E_\al$ and $\psi_\al
(n)$ are the eigenvalues and eigenstates of the ${\hat H}_0$ ($\al$ and $n$ 
label the states and sites respectively), the initial density matrix is given 
by 
\beq
{\hat \rho}_{mn} (0) ~=~ \sum_\al ~\psi_\al (m) ~\psi_\al^* (n) ~f(E_\al , 
\mu ,T) ~.
\label{rho0}
\eeq
We then evolve the density matrix in time and compute the current and charge 
transferred using Eqs. (\ref{rhot}-\ref{dqt}). 

We will also be interested in calculations in which the chemical potentials 
are not the same in the two leads; suppose that they are given by $\mu_L$ and 
$\mu_R$ in the left and right leads respectively. In this case, we begin by 
setting the hopping amplitude to be zero on one of the bonds in the middle of 
the system; we then compute the density matrices $\hat{\rho}_L$ and 
$\hat{\rho}_R$ in the left and right parts of the system by restricting Eq. 
(\ref{rho0}) to the left and right leads respectively. The complete density 
matrix $\hat{\rho} (0)$ is then given by the direct sum of $\hat{\rho}_L$ and
$\hat{\rho}_R$. We then set the hopping amplitude on that bond to unity, and 
evolve the density matrix with the Hamiltonian $\hat{H} (t)$ as usual.

We should note that although the net charge transferred per cycle is the same
across all bonds, the detailed time dependence of the charge transferred looks
different for different bonds. (This will become particularly clear in a later
figure). Another point to note is that the finite length of the leads (with 
$N_L$ sites) implies that the system has a return time $T_R$ equal to $2N_L /
v_F$ where the Fermi velocity $v_F = 2 \ga \sin k_F$ \cite{dhar2}; this is the
time required for an electron to travel from the wire in the middle to the end
of either of the two leads and then return to the wire. The numerical results 
can be trusted only for times which are less than $T_R$. Finally, there are 
transient effects which last for one or two cycles; the effects of different
choices of the initial density matrix get washed out after this transient 
period. All the numerical results presented below are therefore taken from 
times which are larger than the transient time but smaller than $T_R$.

\begin{figure}[htb]
\begin{center}
\epsfig{figure=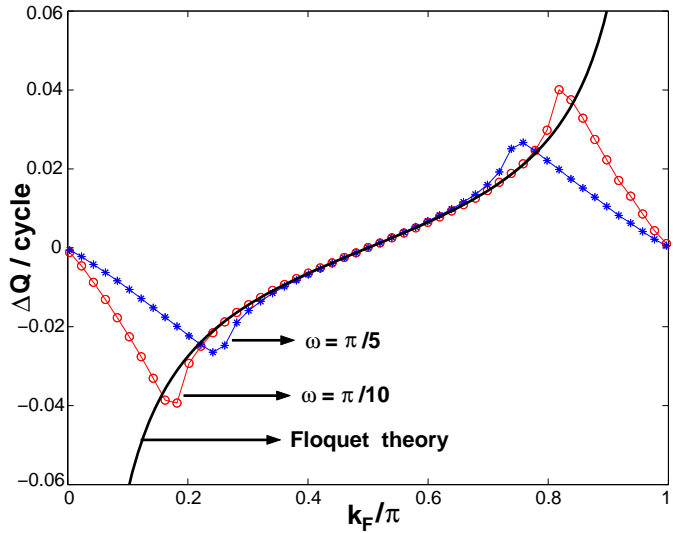,width=9cm} 
\end{center}
\caption{Charge transferred (in units of $q$) per cycle from left to right 
versus the Fermi wavenumber for an oscillating potential at 2 sites, with 
$a_1 = a_2 = 0.2$, $\phi_2 - \phi_1 = \pi /2$, and $\om = \pi /10$ (circles) 
and $\pi /5$ (stars) for a system with 202 sites ($N_L = 100$). The solid line
shows the results obtained using second order Floquet scattering theory in
the limit $\om \to 0$.}
\end{figure}

\section{Numerical results} 

\subsection{Effect of band filling}

Figure 1 shows the charge transferred versus the Fermi wavenumber $k_F$ as 
obtained by the EOM method when oscillating potentials are only applied at two
neighboring sites. If the amplitudes $a_1$ and $a_2$ are much smaller than 
$v_F$, one can use Floquet scattering theory to second order in the $a_n$ 
to find the charge transferred per cycle from left to right
in the limit $\om \to 0$,
\beq
\Delta Q ~=~ -~ q ~\frac{a_1 a_2}{\ga^2} ~\sin (\phi_2 - \phi_1) ~
\frac{\cot k_F}{2} ~.
\label{cur1}
\eeq
The comparison between the results obtained numerically for two different
pumping frequencies and the analytical expression given in Eq. (\ref{cur1}) 
is also shown in Fig. 1. We see that the numerical results and second order 
Floquet scattering theory match near the middle of the band, $k_F = \pi /2$, 
but the agreement becomes poor near the band edges $k_F = 0$ and $\pi$. This
discrepancy is due to the expansion parameter $a_n /v_F$ becoming large near 
the band edges; hence second order perturbation theory breaks down, and Eq. 
(\ref{cur1}) is no longer valid. We can also define an adiabaticity
parameter $\ep = (d/v_F)/(2\pi /\om)$ which is the ratio of the time
taken for an electron with Fermi velocity $v_F$ to traverse the scattering
region of length $d$ to the time period of the oscillating potentials; the 
adiabatic limit corresponds to $\ep \to 0$. In our case, $d=1$ (the
oscillating potentials are at neighboring sites), while 
$v_F =2$ at $k_F = \pi/2$ (half-filling). Even for the largest frequency 
of $\om = \pi /5$ used in Fig. 1, we see that $\ep = 1/20$. This is why 
the two curves with finite $\om$ collapse on to the results of Floquet 
scattering theory near $k_F = \pi/2$; this may not happen if $\ep$ is of 
order 1, i.e., if either $\om$ or $d$ becomes larger.

Fig. 1 shows that the charge transferred goes through an extremum as $k_F$ 
approaches the band edges (0 or $\pi$). This can be understood as follows. For
a weak potential ($a_1, a_2$ small), only the first Floquet side bands ($p = 
\pm 1$) contribute to the current. From the structure given in Eq. (\ref{ir}),
one can then see that at zero temperature, the current gets a contribution only
from electrons which lie within the energy range $[E_F - \om , E_F]$ initially
and get excited to the energy range $[E_F , E_F + \om]$ finally. The number of
states lying within these ranges starts decreasing when one gets very close to
the band edges, namely, when $E_F - \om$ goes below the bottom end of the band,
or $E_F + \om$ goes above the top end. (We recall that the band has a finite 
width going from $-2 \ga$ to $2 \ga$). Hence the magnitude of the current 
starts decreasing when $E_F - \om$ falls below $- 2 \ga$ or when $E_F + \om$ 
goes above $2 \ga$. Qualitatively, this is what one observes in Fig. 1; the 
charge transferred goes through an extremum when $E_F = - 2 \ga k_F$ gets 
within a distance of $\om$ from the top end or the bottom end of the band. 
The extremum occurs closer to the band edge if $\om$ is smaller ($\pi/10$ 
instead of $\pi/5$).

Note that the Hamiltonian in Eqs. (\ref{h0}-\ref{pot}) is invariant under the 
particle-hole transformation $c_n \to (-1)^n c_n^\dag$ and $t \to t + \pi 
/\om$, but the current operator changes sign. Hence the charge transferred is 
antisymmetric about $k_F = \pi /2$ (half-filling) as we can see in Fig. 1;
hence the pumped charge is exactly zero at half-filling for any value of
the frequency $\om$.

\subsection{Traveling potential wave}

Eq. (\ref{cur1}) can be generalized to the case in which there are oscillatory
potentials at several sites as given in Eq. (\ref{pot}). In the limit $\om \to
0$, the use of Floquet scattering theory up to second order in the amplitudes 
gives the charge transferred per cycle from left to right to be
\beq
\Delta Q = -~ q ~\sum_{n>m} ~\frac{a_n a_m}{\ga^2} ~\sin (\phi_n - \phi_m) ~
\frac{\sin [2k_F (n-m)]}{4 \sin^2 k_F} ~.
\label{cur2}
\eeq
Since the amplitudes $a_n$ are positive, Eq. (\ref{cur2}) suggests that the 
charge transferred will be maximized if one chooses $\phi_n - \phi_m$ to be in
phase with $2k_F (n-m)$ for all pairs of sites $n$ and $m$. A simple way to 
ensure this is to choose $\phi_n = 2 k_F n$. This is the choice of the phases
$\phi_n$ made in Figs. 2-6 all of which involve systems with oscillating 
potentials applied to 8 consecutive sites, with the same amplitude $a$ at all
those sites. Hence the potential at site $n$ takes the form $a \cos (\om t + 2
k_F n)$; this describes a potential wave traveling with a velocity $\om /2k_F$.

\begin{figure}[h!]
\begin{center}
\epsfig{figure=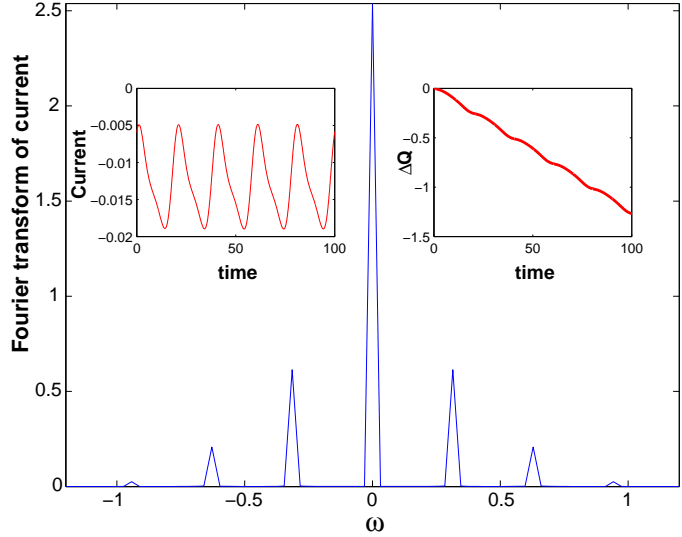,width=9cm}
\end{center}
\caption{Fourier transform of the current (in units of $q$) versus frequency 
for oscillating potentials at 8 sites, with $a=0.2$, $\om = \pi /10$ and $k_F 
= \pi /4$ for a system with 638 sites ($N_L = 315$). The left and right insets
show the current and charge transferred versus time.}
\end{figure}

\begin{figure}[h!]
\begin{center}
\epsfig{figure=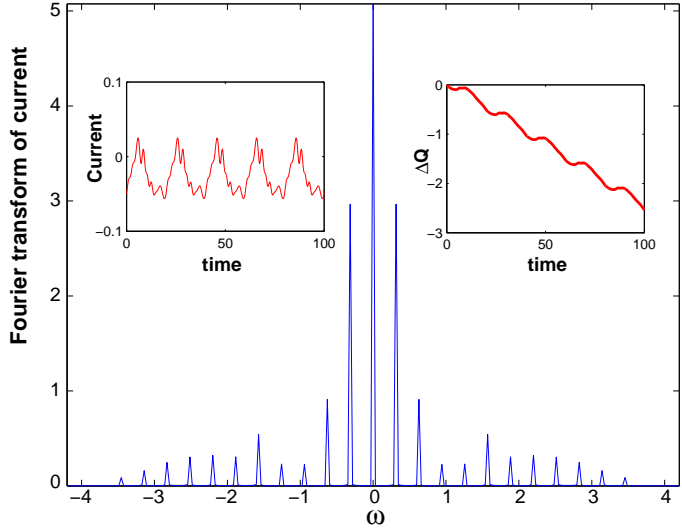,width=9cm} 
\end{center}
\caption{Fourier transform of the current (in units of $q$) versus frequency 
for oscillating potentials at 8 sites, with $a=2$, $\om = \pi /10$ and $k_F 
= \pi /4$ for a system with 638 sites ($N_L = 315$). The left and right insets
show the current and charge transferred versus time.}
\end{figure}

Figures 2 and 3 compare the cases of weak and strong pumping. The Fourier 
transform of the current (computed at the ninth bond to the right of the 
scattering region) shows that for weak pumping ($a=0.2$), only a small number 
of Floquet side bands contribute to the current, while for strong pumping 
($a=2$), a large number of Floquet side bands contribute. The Fourier
transform of the current therefore provides a way of distinguishing between 
strong and weak pumping.

\begin{figure}[h!]
\begin{center}
\epsfig{figure=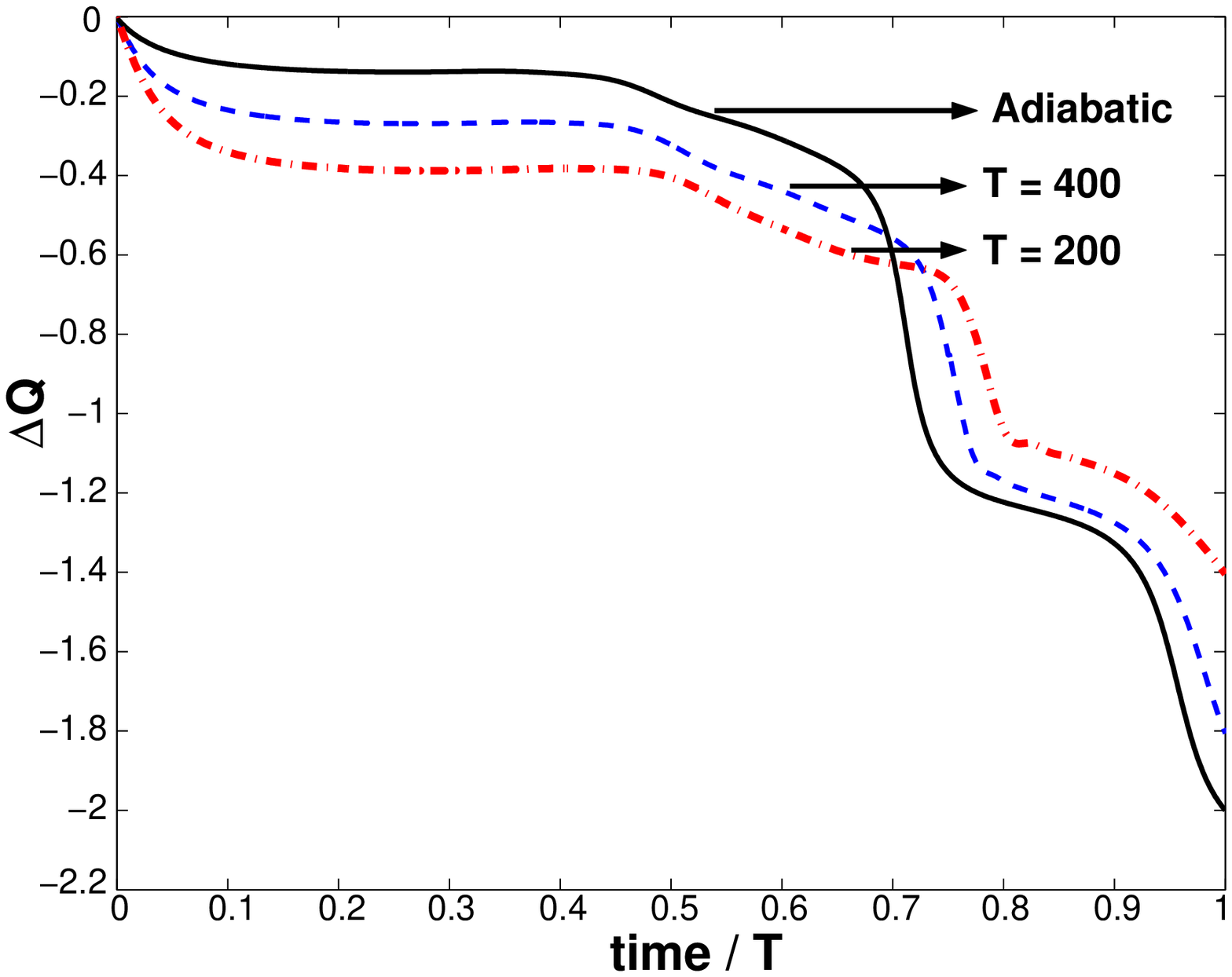,width=9cm} 
\end{center}
\caption{Charge transferred (in units of $q$) from left to right versus time 
during one cycle (with period $T = 2 \pi /\om$) for oscillating potentials at 
8 sites, with $a=2$, $k_F = \pi /8$, and $\om = \pi /100$ (dash dot), 
$\pi /200$ (dashed), and tending to zero (solid line).}
\end{figure}

As an application of the EOM method to recent observations of charge pumping by
a traveling potential wave \cite{aizin,talyanskii}, we present in Fig. 4 the 
charge transferred versus time in one cycle in the strong pumping regime, as 
obtained by the EOM method for $\om = \pi /100$ and $\pi /200$, and from 
Eq. (\ref{btp}) for the adiabatic case. We see that very little charge is 
transferred in one part of the cycle, and a lot of charge is transferred in 
the other part; the reason for this will become clear below. The charge 
transferred per cycle is about $1.4$ for $\om =\pi /100$, $1.8$ for $\om = 
\pi /200$, and exactly 2 in the adiabatic case. In this model, therefore, 
the charge transferred increases as the pumping becomes more adiabatic.

\begin{figure}[h!]
\begin{center}
\epsfig{figure=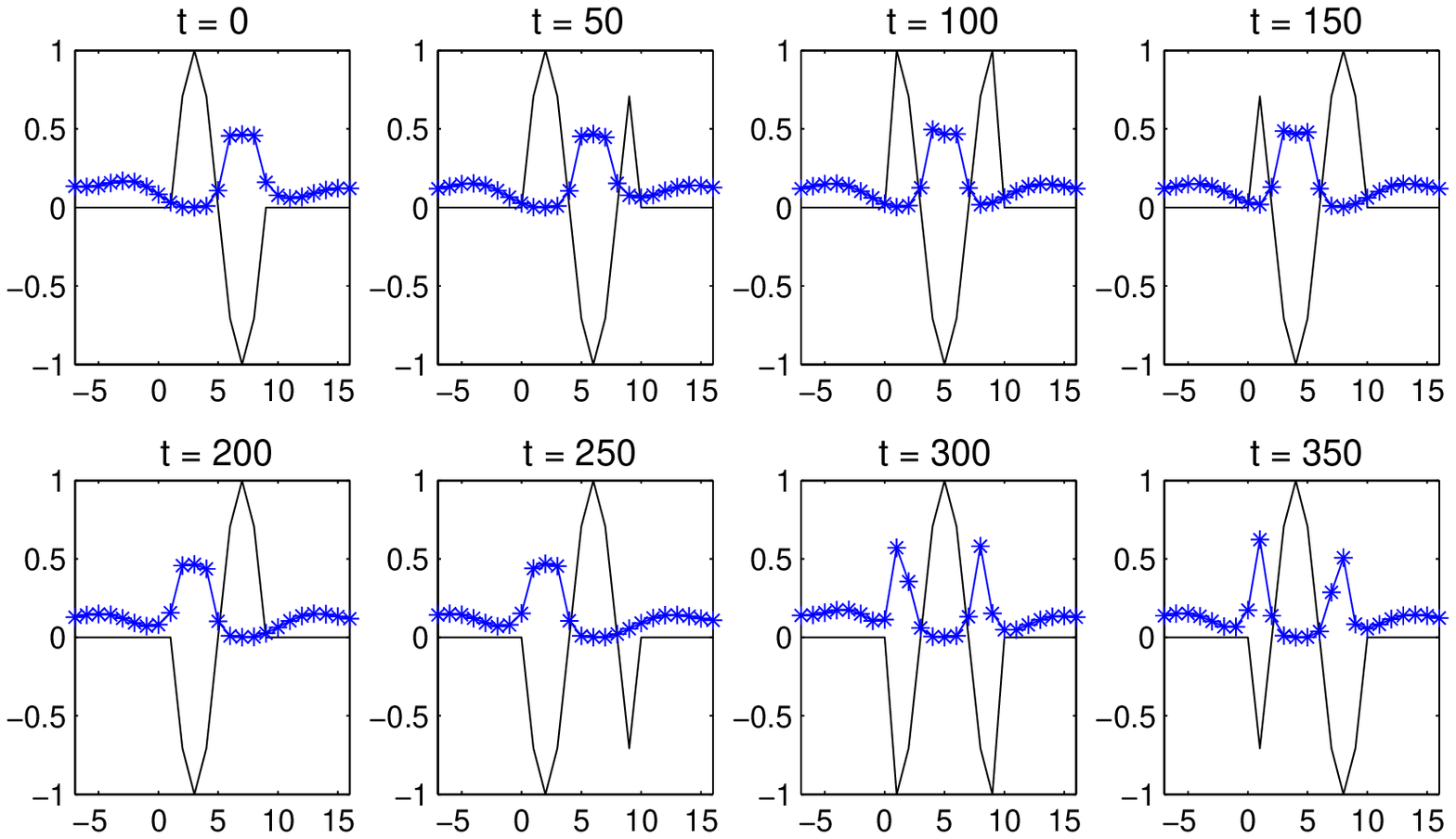,width=12cm} 
\end{center}
\caption{Electron numbers (stars) and potentials at 24 sites in and on two 
sides of the scattering region (sites 1 to 8) at 8 equally spaced times in one
cycle, for potentials oscillating with $a=2$, $\om = \pi /200$ and $k_F = 
\pi /8$ for a system with 1618 sites ($N_L = 805$). The scale on the $y$-axis 
indicates both the electron number and the potential (divided by 2) at each 
site.}
\end{figure}

Figure 5 shows the density profile, $Tr ~(\hat{\rho} ~c_n^\dag c_n)$,
in and on two sides of the scattering region at eight equally spaced times in
one cycle for the same parameters as in Fig. 4, with $\om = \pi /200$. 
We see a larger number of electrons in the regions where the 
potential has a minimum; these electrons move along with the potential 
minimum. The first six pictures in Fig. 5 show some electrons (about $1.8$ in 
number as indicated in Fig. 4) being transported by the potential minimum 
from the right side of the scattering region to the left; during this period,
very little charge is transferred to or from the leads. The last two pictures 
in Fig. 5 show these electrons being transmitted to the left lead, while some 
other electrons are entering the scattering region from the right lead. These 
pictures illustrate the mechanism of charge transfer mentioned in Ref. 
\cite{talyanskii}. Note that charge gets pumped in this model even though the
`frozen' $S$-matrix is almost perfectly reflecting at all times; the current 
would have been very small if the potential wave had been stationary.

\subsection{Asymmetry in charge and energy currents}

Although the time-averaged current must be the same at all sites due to
current conservation, the fluctuations in the current need not be the same 
everywhere. This is illustrated in Figs. 6 and 7 which show the currents at 
the tenth bond on the left and tenth bond on the right respectively of the 
scattering region for a model with the same parameters as in Fig. 3. We see 
that the current is distinctly more noisy on the left. This is an effect of 
non-adiabaticity; we find that it disappears in the limit $\om \to 0$.
Qualitatively, this asymmetry occurs because the potential minimum `picks' 
up electrons from the right lead, transports them through the 
scattering region with a finite velocity given by $\om /2k_F$, and 
finally `throws' them into the left lead with that velocity; the finiteness
of this velocity may be responsible for the additional noise on the left.

\begin{figure}[h!]
\begin{center}
\epsfig{figure=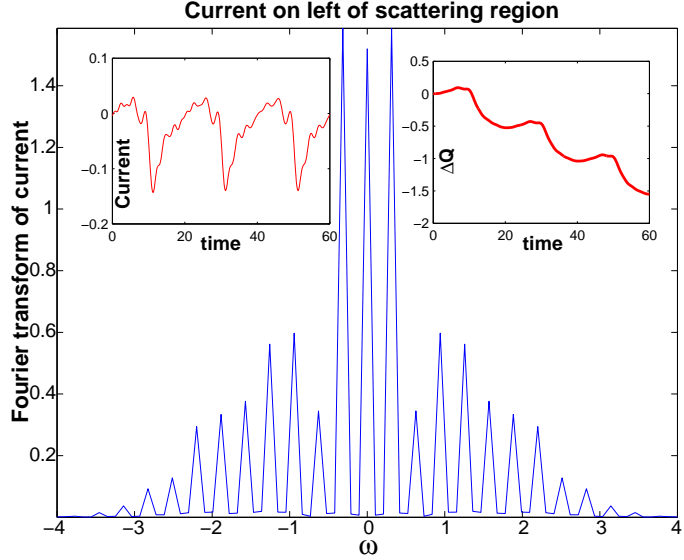,width=9cm} 
\end{center}
\caption{Fourier transform of the current (in units of $q$) on the left of the
scattering region versus frequency for oscillating potentials at 8 sites, with
$a=2$, $\om = \pi /10$ and $k_F = \pi /4$ for a system with 428 sites ($N_L = 
210$). The left and right insets show the current and charge transferred 
versus time.}
\end{figure}

\begin{figure}[h!]
\begin{center}
\epsfig{figure=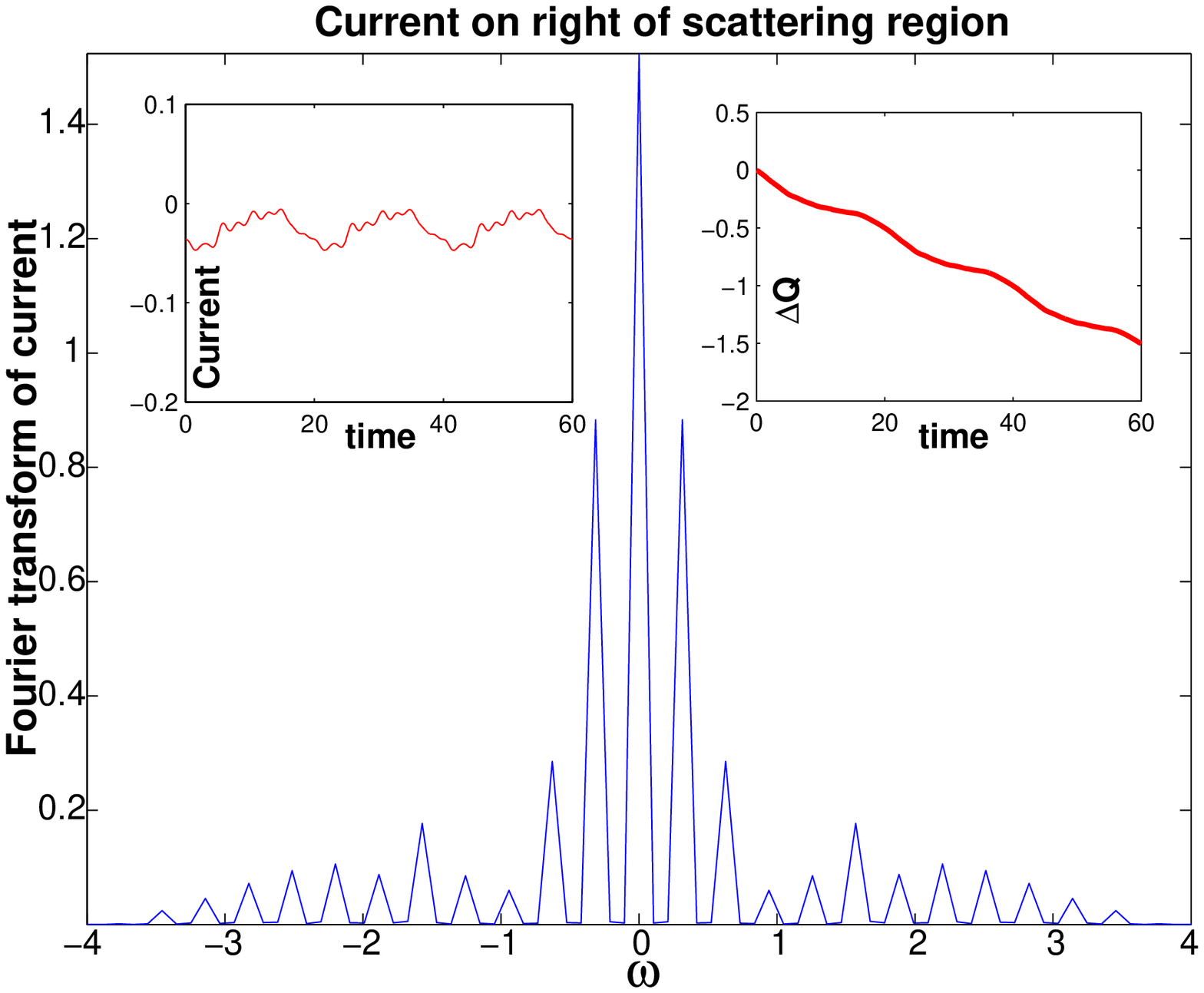,width=9cm} 
\end{center}
\caption{Fourier transform of the current (in units of $q$) on the right of 
the scattering region versus frequency for oscillating potentials at 8 sites, 
with $a=2$, $\om = \pi /10$ and $k_F = \pi /4$ for a system with 428 sites 
($N_L = 210$). The left and right insets show the current and charge 
transferred versus time.}
\end{figure}

\begin{figure}[h!]
\begin{center}
\epsfig{figure=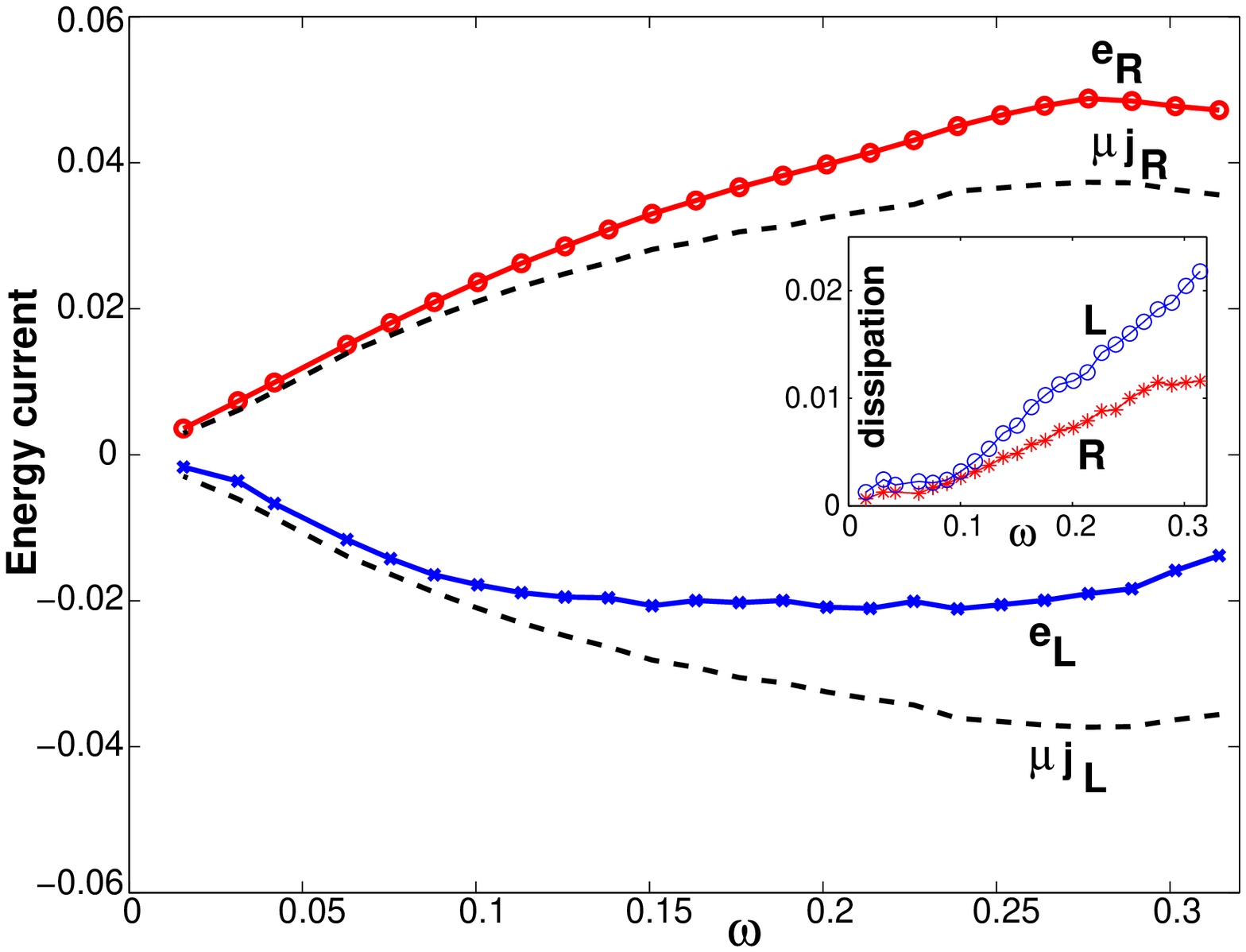,width=9cm} 
\end{center}
\caption{Energy currents $e_L$ and $e_R$ on the left and right of the 
scattering region versus frequency for oscillating potentials at 8 sites, with
$a=2$ and $k_F = \pi /4$. The curves marked as $\mu j_L$ and $\mu j_R$ show 
the product of the chemical potential with the charge currents $j_L$ and $j_R$
on the left and right. The inset shows the heat current on the left and right 
versus frequency; these are given by $e_L - \mu j_L$ and $e_R - \mu j_R$ 
respectively.}
\end{figure}

The asymmetry can also be seen in the energy and heat currents
\cite{avron,moskalets,dhar1} on the left and right of the scattering region 
for the same model. Figure 8 shows the {\it outgoing} energy currents $e_L$
and $e_R$ on the tenth site on the left and tenth site on the right 
respectively of the scattering region, and the product of the chemical 
potential $\mu$ with the outgoing charge currents $j_L$ and $j_R$ on the left 
and right, all as functions of the pumping frequency $\om$. (Due to current 
conservation in the steady state, $j_L = -j_R$). The differences between the 
two currents, namely, $e_L - \mu j_L$ and $e_R - \mu j_R$ give the heat 
currents on the left and right; these are shown in the inset of Fig. 8. We see
that the heat current on the left is significantly larger than on the right. We
have not attempted to determine quantitatively how the heat current varies with
$\om$; the dependence is known to be quadratic for small $\om$ 
\cite{avron,moskalets}.

We should emphasize here that our model has no mechanisms (such as 
electron-phonon scattering) for heat dissipation in the leads. A real system 
will have such mechanisms, and the heat currents calculated above will 
eventually get dissipated somewhere in the leads.

\begin{figure}[h!]
\begin{center}
\epsfig{figure=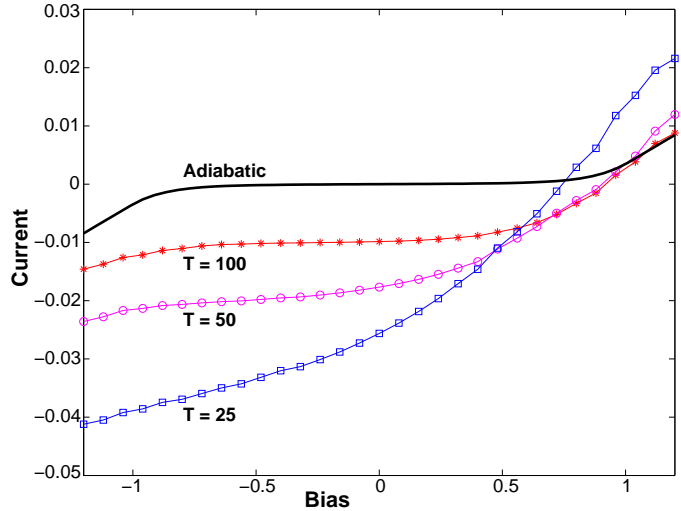,width=9cm} 
\end{center}
\caption{Current (in units of $q$) versus bias for different frequencies (with
$T = 2\pi /\om$) for oscillating potentials at 8 sites, with $a=2$ and $k_F = 
\pi /4$ for a system with 428 sites ($N_L = 210$).}
\end{figure}

\subsection{Effect of finite bias}

We will now consider the effect of a finite bias on the current \cite{entin2}.
We take the chemical potential in the left and right leads to be $\mu_L =
\mu + qV/2$ and $\mu_R = \mu - qV/2$ respectively, so that there is a bias of 
$qV$ between left and right. The calculations are done as outlined in Sec. IV.
Fig. 9 shows the effect of a bias on the current for different 
frequencies for the same model as in Fig. 3. As expected, the current from left
to right increases with the bias; however, the current continues to be negative
(i.e., flows from right to left) for a range of positive values of the bias. 
For a pumping frequency of $\om = 2 \pi /100 \simeq 0.063$, charge can get 
pumped against an opposing bias of almost $0.9$. We also observe a pronounced
asymmetry between positive and negative values of the bias, and the asymmetry 
increases with the pumping frequency. Finally, note that the current is
very small for small bias in the adiabatic limit; this is because the 
scattering region is almost perfectly reflecting, as was mentioned earlier. 
In the adiabatic limit, we have computed the current analytically using
the Landauer-B\"uttiker expression averaged over time \cite{entin2}. Namely,
\beq
I_R ~=~ q ~\int_{-2\ga}^{2\ga} ~\frac{dE}{2\pi} ~<| S_{RL} (E,t) |^2>_t ~
[~ f(E, \mu + qV/2, T) - f(E, \mu - qV/2 , T) ~] ~,
\eeq
where $<| S_{RL} (E,t) |^2>_t$ denotes the average over one time period of the
`frozen' transmission probability from left to right for a given energy $E$.

\section{Discussion}

The EOM method provides a general way of computing the charge, energy and heat
currents as functions of time at different sites. A knowledge of the detailed 
space-time dependence of currents and charges is often useful. For instance, 
our results give an insight into the mechanism of charge pumping by a traveling
potential wave which has been studied experimentally in several systems 
\cite{talyanskii}. Namely, there are more electrons in a region in which the 
potential is attractive; as this region moves with time, so do those electrons.
In this way, electrons are transported across the wire from one lead to 
another. We find that non-adiabatic charge pumping by a traveling potential 
wave produces an appreciable asymmetry in the Fourier transforms of the
charge and heat currents. It would be interesting to experimentally look for 
an asymmetry in the Fourier transform of the current in the different systems 
where charge pumping has been demonstrated.

The EOM approach is quite versatile and makes no assumptions about the ranges 
of the different parameter values; the potentials and pumping frequency may 
be small or large, and the potentials may vary with time in an arbitrary way,
not necessarily simple harmonic or even periodic. The EOM method can also be 
used to study pumping at resonant frequencies \cite{moskalets,strass}.

It may be useful to compare the EOM and NEGF methods here. The EOM method 
requires long leads in order to have a large return time, so that one has a 
reasonable window of time to compute various quantities of interest. It can be
extended to the case of two- and three-dimensional leads \cite{bushong}; 
however the calculations involve significantly larger systems in those cases.
The NEGF formalism involves a self-energy $\Sigma (E)$ which takes the leads 
into account in an exact way; hence large leads do not need to be explicitly 
included in the numerical computations. On the other hand, for time-dependent
problems such as charge pumping, the NEGF formalism must work with Green 
functions which would generally depend on two time arguments through the 
self-energy $\Sigma (t,t')$. In contrast to this, the EOM approach is local in
time, and is therefore simpler to implement numerically. One can work with the
NEGF approach in the frequency domain, but that works well only if the 
potentials vary harmonically in time. The EOM approach would work even for 
potentials which vary with time in an arbitrary way.

It would be useful to extend the EOM method so as to take into account 
interactions between the electrons \cite{bushong}. For small pumping 
frequencies and weak interactions, one can use an adiabatic Hartree-Fock 
approximation; for instance, an on-site Hubbard interaction of the form 
$U c_{n,\ua}^\dag c_{n,\ua} c_{n,\da}^\dag c_{n,\da}$ can be approximated by 
the time-dependent term $U [{\hat \rho}_{n,n,\ua} (t) ~c_{n,\da}^\dag c_{n,\da}
+ {\hat \rho}_{n,n,\da} (t) ~c_{n,\ua}^\dag c_{n,\ua}]$. However, this 
approximation will break down if the interactions are strong, and also if the 
pumping frequency is large (in which case the effective on-site interaction 
may not instantly follow the on-site densities). We would therefore require 
a different technique in such situations. Recently, Tomonaga-Luttinger liquid 
theory has been used to study the effects of oscillating potentials in 
interacting systems \cite{feldman,sharma,makogon,das}.

\section*{Acknowledgments}

A.A. thanks CSIR, India for a Junior Research Fellowship. D.S. thanks S. Datta,
A. Dhar and S. Rao for stimulating discussions. We thank DST, India for 
financial support under projects SR/FST/PSI-022/2000 and SP/S2/M-11/2000.

\end{document}